\documentclass[twoside]{LCWS11}
\usepackage[latin1]{inputenc}
\usepackage[dvips]{graphicx,epsfig,color}
\usepackage{wrapfig,rotating}
\usepackage{amssymb,amsmath,array,wasysym}
\usepackage{cite}
\def\as{\alpha_{s}}

\def\asb{\bar{\alpha}_{s}}
\def\Ltau{L}

\def\asl{\alpha_{s}\Ltau}

\def\d{\hbox{d}}

\begin{document}
\title{
Thrust distribution resummation in $e^{+}e^{-}$ collisions.} 
\author{Pier~Francesco~Monni
\thanks{in collaboration with Thomas~Gehrmann and Gionata~Luisoni. Talk given at the LCWS 2011, Granada, Spain.}
\vspace{.3cm}\\
Institut f\"ur Theoretische Physik, Universit\"at Z\"urich,
Winterthurerstrasse 190,\\CH-8057 Z\"urich, Switzerland
}

\maketitle

\begin{abstract}
In this talk \cite{linktalk} we report on the recent progresses on IR logarithms resummation for the Thrust distribution in $e^{+}e^{-}$ collisions.
Using renormalisation group (RG) evolution in Laplace space, the resummation of
logarithmically enhanced corrections is performed to next-to-next-to-leading logarithmic (NNLL) accuracy. 
To combine the resummed expressions with the fixed-order results, we derive the $\log(R)$-matching and $R$-matching of the NNLL approximation to the fixed-order NNLO distribution.
\end{abstract}

\section{Introduction}
Event-shapes are observables which measure the geometrical properties of energy-momentum flow in a hadronic final state. They have been 
precisely measured over a broad range in energies at electron-positron colliders. The event-shape distributions allow 
for  a detailed probe of the dynamics of QCD and especially for a precise determination of the 
strong coupling constant~$\alpha_s$. 
Owing to their infrared and collinear
safety, they can be computed systematically  in perturbation theory.
The fixed-order description, based on a power series expansion of the distribution 
in the strong coupling constant, is reliable over most of the 
kinematical range of the event-shape. In the dijet limit, which is attained for 
the thrust variable~\cite{thrust} as $T\to 1$, the convergence of the fixed-order expansion 
is spoilt by large logarithmic terms $\log(1-T)$ at each order in the strong coupling constant,
thus it necessitates a resummed description.
During LEP times, precision studies of 
a standard set of six 
event-shapes were based on the combination of fixed-order NLO 
calculations~\cite{Ellis:1980wv, Ellis:1980nc,Kunszt:1980vt,Vermaseren:1980qz,Fabricius:1981sx,Kunszt:1989km,Giele:1991vf,Catani:1996jh} with NLL 
resummation~\cite{CTTW,broadenings,y3}. To avoid the double counting of terms, both 
expansions need to be matched to each other according to matching procedures such as the $R$ and ${\rm log}(R)$ schemes ~\cite{Jones:2003yv}.
In the recent past, substantial progress was made both on the fixed-order and the 
resummed description of event-shapes. Following the development of new 
methods for calculations of QCD jet observables at NNLO~\cite{ourant}, 
the NNLO corrections to $e^+e^-\to 3$ jets and related event-shape observables 
were computed~\cite{GehrmannDeRidder:2007jk,GehrmannDeRidder:2007hr,GehrmannDeRidder:2007bj,Weinzierl:2009nz,Weinzierl:2009ms,weinzierljetnew}.
More recently, in the context of Soft-Collinear-Effective-Theory, the resummation for 
thrust~\cite{Becher:2008cf,Abbate:2010xh} and the heavy jet mass~\cite{Chien:2010kc} beyond NLL
has been performed and applied for a 
precise determination of $\alpha_s$, and the framework for the resummation of the jet broadening 
distributions has been outlined~\cite{Chiu:2011qc, Becher:2011pf}.
In these calculations, the $\mathcal{O}(\alpha_{s}^2)$ soft corrections were determined only up to a constant term by exploiting the renormalisation 
group invariance of the cross section.
Such term is also needed to unambiguously match the resummed distribution to the NNLO result in the $R$ scheme.
In this talk we report on the direct computation of these corrections and we provide a new resummed formula. Finally we match the latter to the existing 
NNLO prediction comparing two different matching schemes.

\section{Fixed-order and resummed distributions}
The differential thrust distribution in perturbation theory is numerically known at NNLO~\cite{GehrmannDeRidder:2007hr,Weinzierl:2009ms}. At a centre-of-mass energy $Q$ and for
a renormalisation scale $\mu$ it reads
\begin{align}
\frac{1}{\sigma}\, \frac{\d\sigma}{\d \tau}(\tau,Q) &= \bar\alpha_s (\mu) \frac{\d A}{\d \tau}(\tau)+ \bar\alpha_s^2 (\mu) \frac{\d B}{\d \tau} (\tau,x_\mu) \nonumber\\
\label{eq:fixedordercs0}
&+ \bar\alpha_s^3 (\mu) \frac{\d C}{\d
\tau}(\tau,x_\mu) +
{\cal O}(\bar\alpha_s^4)\;,
\end{align}
where we defined
\begin{center}
\begin{align}
\asb = \frac{\alpha_s}{2\pi}\;, \qquad x_\mu = \frac{\mu}{Q}\;,
\end{align}
\end{center}
and where $\sigma$ is the total perturbative hadronic cross-section for $e^{+}e^{-}\rightarrow$ hadrons.
The explicit dependence on the renormalisation scale is given by
\begin{align}\label{eq:fixedordermudep}
\frac{\d B}{\d \tau}(\tau,x_\mu) &=\,\frac{\d B}{\d \tau}(\tau)+2\beta_{0}\,\log(x_{\mu}^{2})\frac{\d A}{\d \tau}(\tau),\\
\frac{\d C}{\d \tau}(\tau,x_\mu) &=\,\frac{\d C}{\d \tau}(\tau)+2\,\log(x_{\mu}^{2})\big(2\beta_{0}\frac{\d B}{\d \tau}(\tau) \nonumber\\
&+2\beta_{1}\frac{\d A}{\d\tau}(\tau)\big)+\left(2\beta_{0}\,\log(x_{\mu}^{2})\right)^{2}\frac{\d A}{\d \tau}(\tau).
\end{align}
The QCD $\beta$-function is defined by the renormalisation group equation for the QCD coupling constant 
\begin{align}
\label{rgealpha}
\frac{d\alpha_{s}(\mu)}{d\log\mu^{2}}=-\alpha_{s}(\mu)\bigg(\frac{\alpha_{s}(\mu)}{\pi}\beta_{0}+\frac{\alpha_{s}^{2}(\mu)}{\pi^{2}}\beta_{1}+\ldots \bigg).
\end{align}
The normalised thrust cross-section is then defined as
\begin{align}\label{eq:Rfixed}
R_{T}(\tau)\,\equiv\,\frac{1}{\sigma}\int_{0}^{1}\frac{d\sigma\left(\tau',Q\right)}{d\tau'}\Theta(\tau-\tau^{\prime})d\tau',
\end{align}
where $\sigma$ is the total cross section for $e^{+}e^{-}\rightarrow$ hadrons.
In the two-jet region the fixed-order thrust distribution is enhanced by large infrared logarithms which spoil the convergence of the perturbative series. The convergence can be restored by resumming
the logarithms to all orders in the coupling constant.  The matched cross section can in general be written as
\begin{align}\label{eq:Rres}
R_{T}(\tau) = C(\alpha_{s}) \Sigma(\tau,\alpha_{s})+D(\tau,\alpha_{s}),
\end{align}
where
\begin{align}
\label{eq:csigma1}
C(\alpha_{s}) &= 1+\sum_{k=1}^{\infty}C_{k}\asb^{k},\\
\log\Sigma(\tau,\alpha_{s}) &= \sum_{n=1}^{\infty}\sum_{m=1}^{n+1}G_{nm}\bar{\alpha_{s}}^{n}L^{m}\nonumber\\
\label{eq:csigma2}
=\Ltau g_{1}(\asl)&+g_{2}(\asl)+\frac{\alpha_{s}}{\pi}\beta_{0}g_{3}(\asl)+\ldots
\end{align}

\noindent where $L\equiv\log(1/\tau)$. The function $g_{1}$ encodes all the leading logarithms, the function $g_{2}$ resums all next-to-leading logarithms and so on.
The constant terms $C_{i}$ are required to achieve a full N$^{1+i}$LL accuracy.
$D(\tau,\alpha_{s})$ is a remainder function that vanishes order-by-order in perturbation theory
in the dijet limit $\tau\rightarrow 0$.

 In view of matching the NNLL resummed distribution to the NNLO fixed 
order prediction using the $R$-matching scheme, we need to include the logarithmically subleading terms $C_{2}$, $C_{3}$ 
and $G_{31}$ in the expansions (\ref{eq:csigma1}),(\ref{eq:csigma2}).

The resummation of the thrust distribution beyond NLL was first achieved in \cite{Becher:2008cf} using an effective-theory approach and revisited in \cite{Monni:2011gb},
where the full analytic expressions for the $\mathcal{O}(\bar{\alpha}_{s}^2)$ constant term $C_{2}$ and the coefficient $G_{31}$ were derived.
The $\mathcal{O}(\bar{\alpha}_{s}^3)$ constant term $C_{3}$ is currently unknown, and a numerical estimate is given in \cite{Monni:2011gb} together with the full analytic expressions of 
the functions $g_{i}(\alpha_{s}L)$.\\

\subsection{Factorisation and Resummation}
\begin{figure}[!htp]
\begin{center}
\includegraphics[width=80mm]{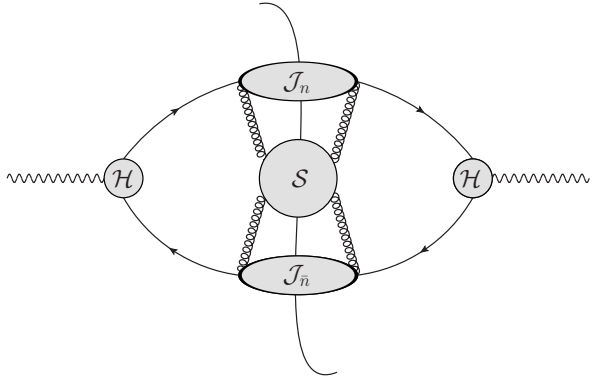}
\caption{Leading regions in dijet factorisation.\label{fig:factorization}}
\end{center}
\end{figure}

Factorisation properties of event-shapes have been widely studied in the 
literature~\cite{CollinsSoper,BerSterKucs1,hoangtop}. Referring to Fig. \ref{fig:factorization} we recast the cross section (\ref{eq:Rfixed})
as
\begin{align}
\label{factorization1}
R_{T}(\tau)=\,\,H\left(\frac{Q}{\mu},\alpha_{s}(\mu)\right)&\int dk^{2}d\bar{k}^{2}\mathcal{J}\left(\frac{k}{\mu},\alpha_{s}(\mu)\right)\mathcal{\bar{J}}\left(\frac{\bar{k}}{\mu},\alpha_{s}(\mu)\right)\notag\\
\times&\int dw\mathcal{S}\left(\frac{w}{\mu},\alpha_{s}(\mu)\right)\Theta(Q^{2}\tau-\bar{k}^{2}-k^{2}-wQ)+\mathcal{O}(\tau),
\end{align}
where we neglected terms of order $\mathcal{O}(\tau)$ which are absorbed in the remainder function $D(\tau,\alpha_{s})$.
We use the integral representation of the $\Theta$-function
\begin{align}
 \Theta(Q^{2}\tau-\bar{k}^{2}-k^{2}-wQ)=\frac{1}{2\pi i}\int_{C} \frac{d\nu}{\nu}\,{\rm e}^{\nu\tau Q^{2}}{\rm e}^{-\nu k^{2}}{\rm e}^{-\nu \bar{k}^{2}}{\rm e}^{-\nu wQ},
\end{align}
and the Laplace transform to recast Eq.~(\ref{factorization1}) as
\begin{align}
R_{T}(\tau) =& H\left(\frac{Q}{\mu},\alpha_{s}(\mu)\right)\frac{1}{2\pi i}\int_{C} \frac{dN}{N}\,{\rm e}^{\tau N}
\tilde{J}^{2}\left(\sqrt{\frac{N_{0}}{N}}\frac{Q}{\mu},\alpha_{s}(\mu)\right)
\tilde{S}\left(\frac{N_{0}}{N}\frac{Q}{\mu},\alpha_{s}(\mu)\right)
\label{eq:RT}
\end{align}
where we set $N=\nu Q^{2}$ and $N_{0}=e^{-\gamma_{E}}$. 
The soft subprocess $\tilde{S}\left({N_{0}}/{N}{Q}/{\mu},\alpha_{s}(\mu)\right)$ describes the interaction between the two jets of hard collinear particles through soft gluon exchange. 
It can be therefore defined in a gauge invariant way as a correlator of Wilson lines
\begin{align}
\tilde{S}\left(\frac{N_{0}}{N}\frac{Q}{\mu},\alpha_{s}(\mu)\right) = \frac{Q}{N_{c}}\int d\tau_{s}{\rm e}^{-\tau_{s}N}\sum_{k_{eik}}\langle0|W^{\dagger}_{\bar{n}}(0)W^{\dagger}_{n}(0)|k_{eik}\rangle
\mathcal{J}_{cut}(\tau_{s}Q)\langle k_{eik}|W_{n}(0)W_{\bar{n}}(0)|0\rangle,
\label{eq:softsub}
\end{align}
 where we defined $\tau_{s}=w/Q$. $W_{n}$ and $W_{\bar{n}}$ are Wilson lines 
\begin{align}
\label{eq:wilsonline}
W_{n}(y) = {\rm \textbf{P}}{\rm exp}\left(ig\int_{0}^{\infty}ds\,n\cdot A(ns+y)\right),
\end{align}
describing the eikonal interaction of soft gluons with the fast moving quarks along the light-like directions $n^{\mu}$ and $\bar{n}^{\mu}$ respectively. $A(ns+y)$ in eq. (\ref{eq:wilsonline}) 
denotes the gluon field
in QCD. The sum runs over the final states $|k_{eik}\rangle$ involving $k$ soft particles whose phase space is constrained according to the thrust trigger function $\mathcal{J}_{cut}(\tau Q^{2})$. 
Both soft and soft-collinear contributions are encoded into the soft subprocess. The two-loop expression was computed analytically in \cite{Monni:2011gb} by performing direct phase-space cuts.
The results are in agreement with those presented in \cite{schwartznew, stewartnew}.
The collinear subprocess $\mathcal{J}$ ($\mathcal{\bar{J}}$) describes the decay of the jet-initiating hard quark (antiquark) into a jet of collinear particles moving along the $n^{\mu}$ ($\bar{n}^{\mu}$)
direction. It is therefore an inclusive quantity which can be found in many other relevant QCD processes such as deep inelastic scattering and heavy quarks decay \cite{Ster87,KorSter,SCETJet}.
The short-distance hard function $H\left({Q}/{\mu},\alpha_{s}(\mu)\right)=|\mathcal{H}\left({Q}/{\mu},\alpha_{s}(\mu)\right)|^{2}$ takes into account the hard virtual corrections to 
the quark-antiquark production subprocess. It is free of large logarithms and it can be generally defined such that Eq. (\ref{factorization1}) reproduces the fixed-order cross section up to
power suppressed terms.\\
Using the Renormalisation Group evolution of the soft and collinear subprocesses \cite{KorSter,SCETJet,Monni:2011gb}, we can recast eq.~(\ref{eq:RT}) as
\begin{align}
R_{T}(\tau) =& H\left(\frac{Q}{\mu},\alpha_{s}(\mu)\right)\frac{1}{2\pi i}\int_{C} \frac{dN}{N}\,{\rm e}^{\tau N}
\tilde{J}^{2}\left(1,\alpha_{s}(\sqrt{\frac{N_{0}}{N}}Q)\right)\tilde{S}\left(1,\alpha_{s}(\frac{N_{0}Q}{N})\right)\times\notag\\
\label{resummedcross}
&{\rm exp}\bigg\{-2\int_{\frac{N_{0}}{N}}^{1}\frac{du}{u}\left[\int_{u^{2}Q^{2}}^{uQ^{2}}\frac{dk^{2}}{k^{2}}\mathcal{A}(\alpha_{s}(k^{2}))
+\mathcal{B}(\alpha_{s}(uQ^{2})) \right]\bigg\},
\end{align}
where the two coefficients $\mathcal{A}(\alpha_{s})$ and $\mathcal{B}(\alpha_{s})$ can be computed in perturbation theory. The coefficient $\mathcal{A}(\alpha_{s})$ reads
\begin{align}
\mathcal{A}(\alpha_{s}) &= \Gamma_{\rm{cusp}}(\alpha_{s})-\beta(\alpha_{s})\frac{\partial \Gamma_{\rm{soft}}(\alpha_{s})}{\partial\alpha_{s}},\notag \\
\end{align}
where $\Gamma_{\rm{cusp}}(\alpha_{s})$ and $\Gamma_{\rm{soft}}(\alpha_{s})$ are the cusp and the soft anomalous dimensions respectively.
The former, together with the coefficient $\mathcal{B}(\alpha_{s})$ can be extracted from the asymptotic limit of the $P_{qq}(\alpha_{s},z)$ splitting 
function \cite{korchdis,bechneubdis} as $z\rightarrow 1$ 
\begin{align}
\label{splittingfunction}
 P_{qq}(\alpha_{s},z) \rightarrow 2\frac{\Gamma_{\rm{cusp}}(\alpha_{s})}{(1-z)_{+}}+2\mathcal{B}(\alpha_{s})\delta(1-z)+...
\end{align}
The integration countour in eq. (\ref{resummedcross})
runs parallel to the imaginary axis on the right of all singularities of the integrand. 
From eq. (\ref{resummedcross}) we see that the $u$-integral in the exponent is regularised by the lower bound $\frac{N_{0}}{N}$. Such a bound acts as
an infrared regulator which prevents the strong coupling constant from being evaluated at non-perturbative scales ($\leq \Lambda_{QCD}$). Then, the contour in eq. (\ref{resummedcross})
should be set away from all the singularities (in particular from the Landau pole). Nevertheless, for resummation purposes we can set the contour on the left
of the Landau singularity since it would contribute with a non-logarithmic effect suppressed with some negative power of the center-of-mass energy scale.
The inversion of the Laplace transform can be performed analytically by using the residue theorem as shown in \cite{CTTW,Monni:2011gb} and results in
\begin{align}\label{eq:Rresummed}
R_{T}(\tau) = \bigg(1+\sum_{k=1}^{\infty}C_{k}\bigg(\frac{\alpha_{s}}{2\pi}\bigg)^{k}\bigg){\rm
exp}\left[\log\frac{1}{\tau}g_{1}(\lambda)+g_{2}(\lambda)+\frac{\alpha_{s}}{\pi}\beta_{0}g_{3}(\lambda)+\left(\frac{\alpha_{s}}{2\pi}\right)^{3}G_{31}\log\frac{1}{\tau}\right],
\end{align}
where 
\begin{align}\label{eq:g_is}
g_{1}(\lambda) =& \,\,\,f_{1}(\lambda),\notag\\
g_{2}(\lambda) =& \,\,\,f_{2}(\lambda)-\log\Gamma(1-f_{1}(\lambda)-\lambda f_{1}^{\prime}(\lambda)),\notag\\
g_{3}(\lambda) =& \,\,\,f_{3}(\lambda)+\left(f_{1}^{\prime}(\lambda)+\frac{1}{2}\lambda
f_{1}^{\prime\prime}(\lambda)\right)\left(\psi^{(0)}(1-\gamma(\lambda))^{2}-\psi^{(1)}(1-\gamma(\lambda))\right)\notag\\
&+f_{2}^{\prime}(\lambda)\psi^{(0)}(1-\gamma(\lambda))+C_{F}/\beta_{0}\left(\gamma_{E}\left(3/2-\gamma_{E}\right)-\pi^{2}/6\right).
\end{align}
The functions $f_{i}(\lambda)$ as well as the constants $C_{1}$, $C_{2}$ and $G_{31}$ are defined in \cite{Monni:2011gb}, while the $C_{3}$ constant term is still analytically unknown.
We fit the latter numerically using the fixed order Monte Carlo parton-level generator {\tt EERAD3}. The fit is performed by subtracting the $\mathcal{O}(\alpha_{s}^3)$ logarithmic structure from
the fixed-order result and taking (numerically) the asymptotic limit $\tau\rightarrow 0$.\\
{\tt EERAD3} is run with a technical cutoff $y_{0}=10^{-5}$ which
affects the thrust distribution below $\tau_{0}\sim \sqrt{y_{0}}$. This forbids us from probing
the far infrared region and we perform the fit for values of $\tau$ larger than $\tau_{0}$.
Numerical fixed order results are obtained with $6\times10^7$ points for the leading colour contribution and $10^7$ points for 
the subleading colour structures.
Because of the presence of large fluctuations in the Monte Carlo results, each color
contribution is fitted separately over an interval where the distribution is
stable and the different results
are combined to find the numerical value of $C_{3}$. As an alternative approach we first sum up all the color contributions and then fit $C_{3}$.
We consider the difference between the two approaches as a systematic error and as
final result we obtain
\begin{equation}
C_{3}=-1050 \pm 180 (\textrm{stat.}) \pm 500 (\textrm{syst.})\,.
\end{equation}

\noindent Considering that there is no statistical correlation between different bin errors, as a different

\begin{wrapfigure}{r}{0.5\columnwidth}
\centerline{\includegraphics[width=0.47\columnwidth]{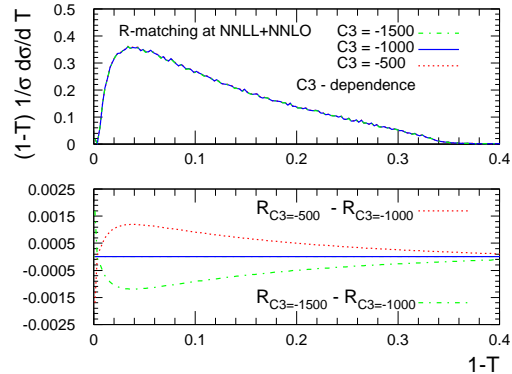}}
\caption{Impact of $C_{3}$ variation}\label{Fig:c3variation}
\end{wrapfigure}

\noindent possible estimate of the systematic 
uncertainty due to the sizeable fluctuation, 
we varied the fit range observing that it does not alter the result in any  
significant way outside the quoted systematic error margins. 
In Fig.~\ref{Fig:c3variation} we vary the value of $C_{3}$ within its error band and we study its impact on the distribution.
We observe that the numerical impact of $C_{3}$ on the distributions is less than $1.5\permil$ and it is
therefore completely negligible compared the other theoretical uncertainties 
such that the large relative error range is tolerable for all practical purposes.
\\
\subsection{Matching to fixed-order and numerical results}
\label{sec:results}
In this section we match the obtained resummed distribution (\ref{eq:Rresummed}) to the NNLO fixed order prediction.
The matching formalism must avoid double counting and allow to access theoretical uncertainties.
We compare the $R$-matching and ${\rm log}(R)$-matching scheme described in \cite{Jones:2003yv}.
\begin{figure}[tbh]
 \centering
 \includegraphics[width=0.5\textwidth]{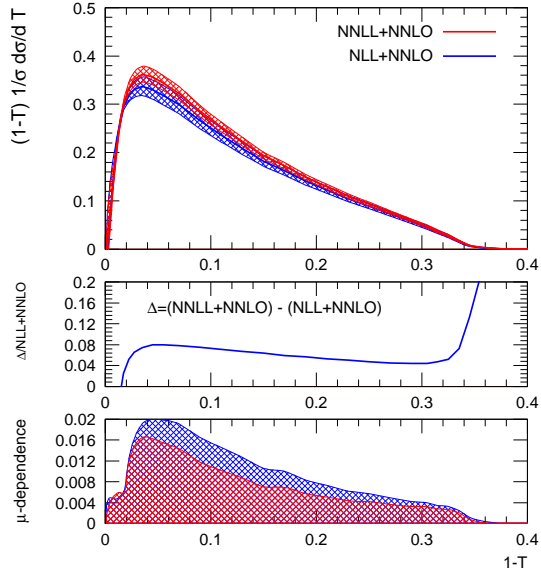}
 \caption{Comparison of the Thrust distributions with
NNLL+NNLO and NLL+NNLO accuracy. The plot on the top shows the two distributions, with the
uncertainty band due to scale
dependence. The curve in the middle shows the difference between NNLL+NNLO and
NLL+NNLO normalised to the NLL+NNLO curve. The impact of the resummation at NNLL is
an increase in the distribution of
order 5-8\%. The lowest plot shows the absolute scale dependence of the two
curves.}
 \label{fig:logrmatching}
\end{figure}

\begin{figure}[htp!]
  \begin{minipage}[c]{.47\textwidth}
    \centering
    \includegraphics[width=1.0\textwidth]{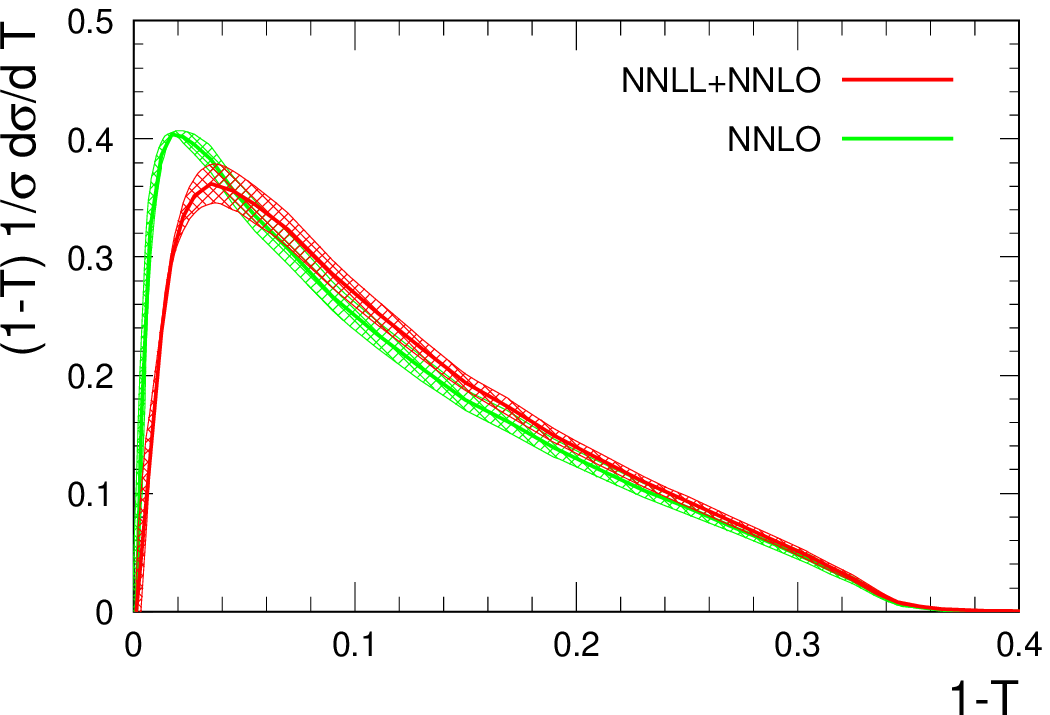}
    \caption{Comparison of the Thrust distribution at NNLO with the matched
NNLL+NNLO predictions. The contribution of NNLL resummation is sizable over the
full thrust range.}
    \label{fig:nnllnnlo}
 \end{minipage}
 \hspace{5mm}
 \begin{minipage}[c]{.47\textwidth}
    \centering
     \includegraphics[width=1.0\textwidth]{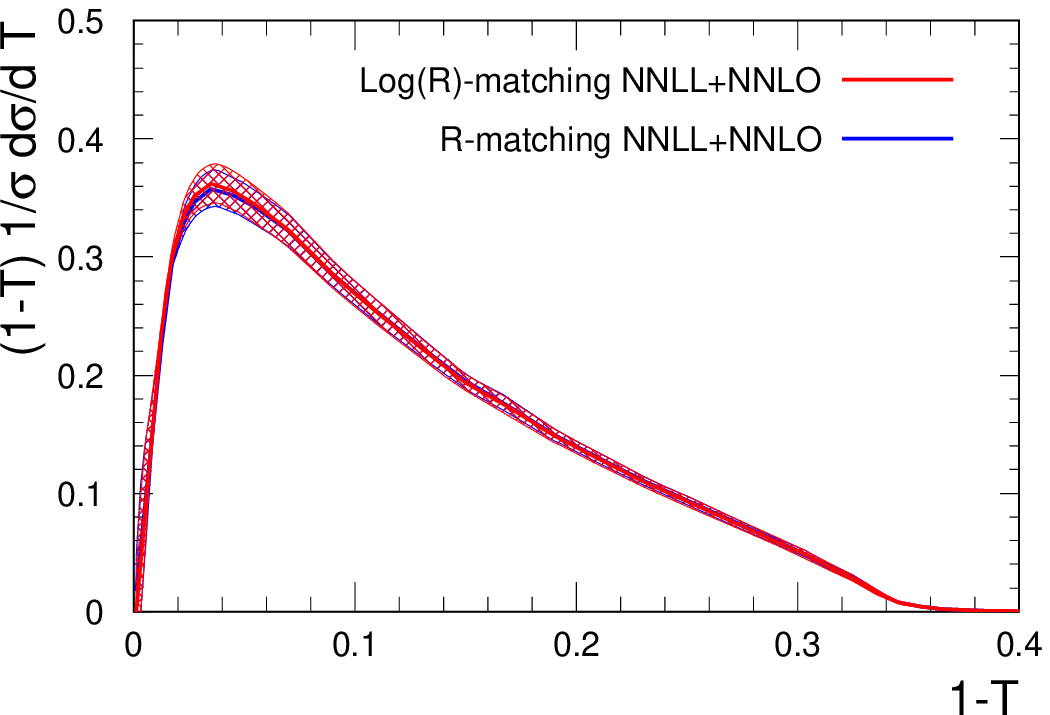}
    \caption{Comparison of the results obtained with the $R$-matching scheme and the
$\log(R)$-matching scheme . The width of the curve shows the uncertainty
related to the scale variation. }
    \label{fig:rmatching}
  \end{minipage}
\end{figure}

In Figure~\ref{fig:logrmatching} we compare the differential cross section of the new
matched NNLL+NNLO results with the old NLL+NNLO derived in~\cite{Gehrmann:2008kh}.
The modification
due to the resummation is sizable, leading to a $8\%$ increase of the distribution
around the peak region. The effect of the additional resummed subleading logarithms
becomes progressively less
important towards the multijet region, where the increase is nevertheless of about
$5\%$. It is interesting to note that the matching of NNLO with NNLL resummation
shifts the pure NNLO result also in
the multijet region (Figure~\ref{fig:nnllnnlo}). This was not the case for NLL+NNLO,
for which the impact of resummation in the region of large $\tau$ was negligible.
This is another sign of the
importance of the NNLL contribution. 
\begin{figure}[htp!]
 \centering
 \includegraphics[width=0.48\textwidth]{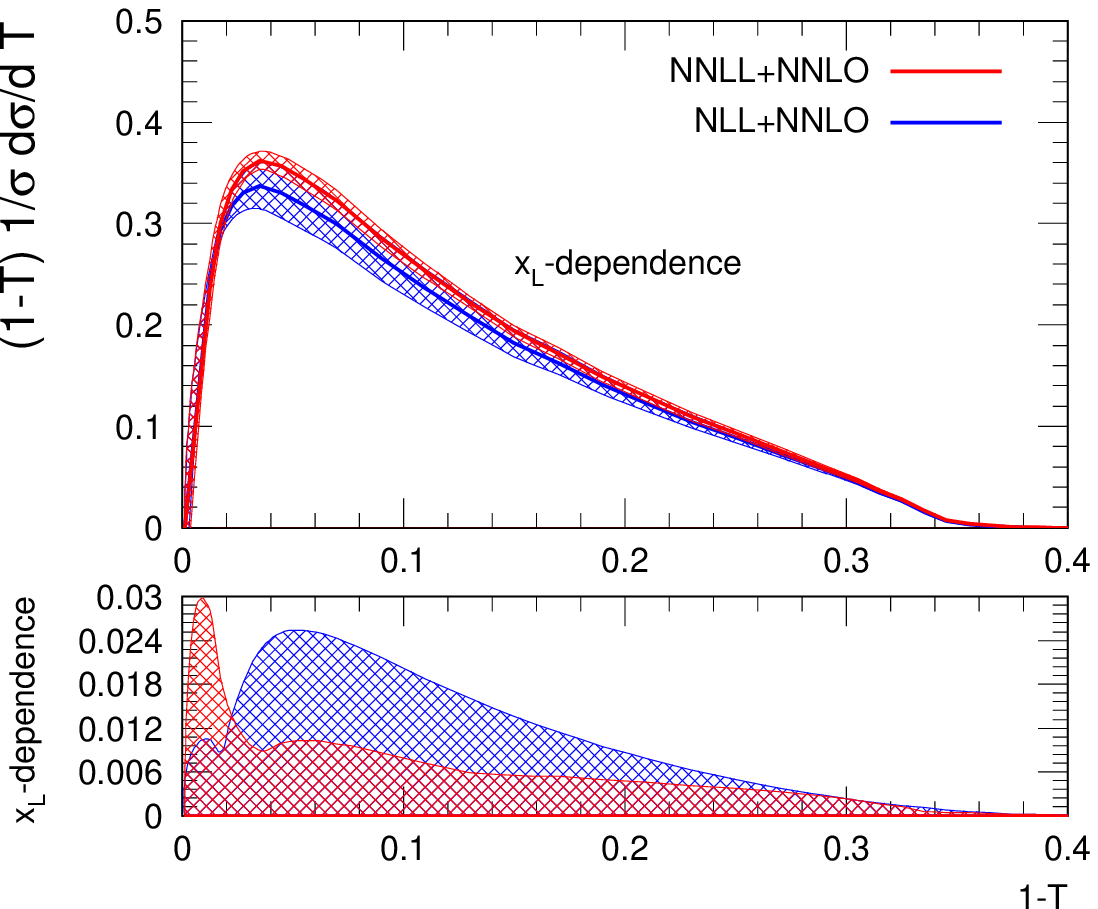}
 \quad
 \includegraphics[width=0.48\textwidth]{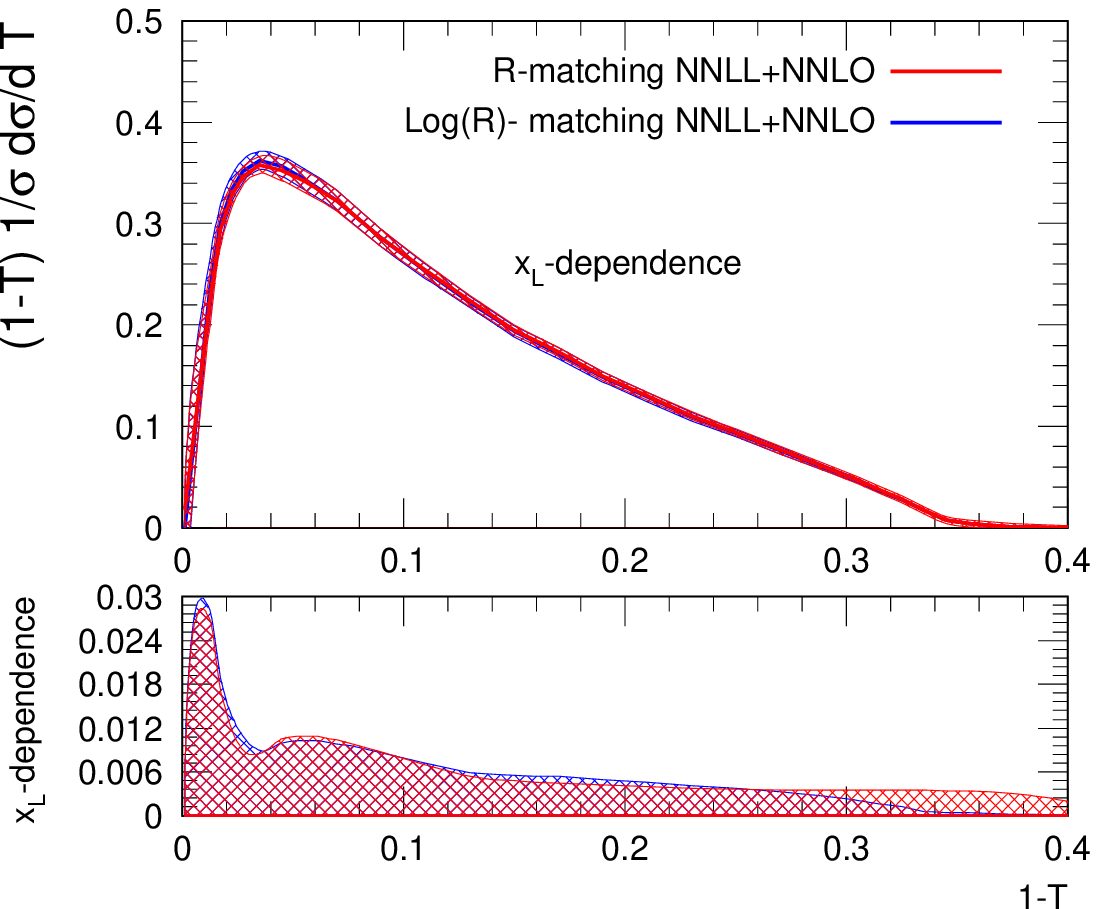}
 \caption{Dependence on the resummed logarithms, determined by varying the parameter
$x_{L}$. The left plot shows the change in the $x_{L}$ dependence between NLL+NNLO
and NNLL+NNLO. The upper plot
shows the distributions with the corresponding uncertainty band, in the lower plot
we compare only the uncertainties. In the right plot the $x_{L}$ dependence using
the two different matching
schemes is shown.}
 \label{fig:matchingxL}
\end{figure}

The renormalisation scale dependence, which was observed to increase from pure NNLO
to NLL+NNLO~\cite{Gehrmann:2008kh,Dissertori:2009ik} because of a mismatch in the
cancellation of renormalisation
scale logarithms, is obtained by varying $0.5<x_{\mu}<2$. It decreases at NNLL+NNLO
by 20\% in the peak region compared to NLL+NNLO. The magnitude of the scale
uncertainty varies between 4\%
in the 3-jet region and 5\% around the peak. 
In Figure~\ref{fig:rmatching} we compare the $R$-matching
and the $\log(R)$-matching
scheme predictions at NNLL+NNLO. The difference between the two matching prescriptions is
tiny and lies well below the scale uncertainty. This implies a very good
stability of the theoretical
predictions under variation of the matching scheme.

One further source of arbitrariness is the choice of the logarithms to be resummed.
In fact, it is not clear whether powers of $\as\log(1/\tau)$ or powers of
{\it e.g.} $\as\log(2/\tau)$ have to be resummed.
The origin of this arbitrariness has to do with how much of the non-logarithmic part
of the fixed-order prediction is exponentiated together with the logarithms. We can
express this arbitrariness by
introducing a new parameter $x_{L}$, which rescales the logarithms as
~\cite{Jones:2003yv}: $L\,\rightarrow\,\hat{L}=\log\left(1/(x_{L}\tau)\right)$.

We can estimate the related uncertainty by varying
the parameter $x_{L}$. In
Ref.~\cite{Jones:2003yv} several prescriptions are given on how to set the correct
variation range for $x_{L}$ for different observables. For the sake of simplicity
and since we are not performing a
fit of the strong coupling constant, we choose to vary $x_{L}$
within the canonical interval $0.5<x_{L}<2$, 
similarly to what is chosen to quantify the
renormalisation scale uncertainty. This
choice is also close to the nominal range of variation proposed
in~\cite{Jones:2003yv}. The impact of this variation is shown in
Figure~\ref{fig:matchingxL}. The left plots show a comparison of the
$x_{L}$-dependence between NLL+NNLO and NNLL+NNLO predictions. The lower plot
allows to quantify the reduction of the uncertainty due to a variation $x_{L}$. 
Apart from the far infrared
region, it is observed to decrease by $50\%$ in the peak region. The scale-dependence reduction is smaller towards the
multijet region, where the contribution of the logarithmic part becomes less
important. The resummation uncertainty at
NNLL+NNLO varies between $2\%$ and $3\%$. In the right plot the same comparison is
made at NNLL+NNLO using the $R$-matching and 
$\log(R)$-matching schemes. We observe a similar $x_{L}$-dependence in
both schemes.

\section{Outlook}
\begin{figure}[htp!]
  \begin{minipage}[c]{.47\textwidth}
    \centering
    \includegraphics[width=1.0\textwidth]{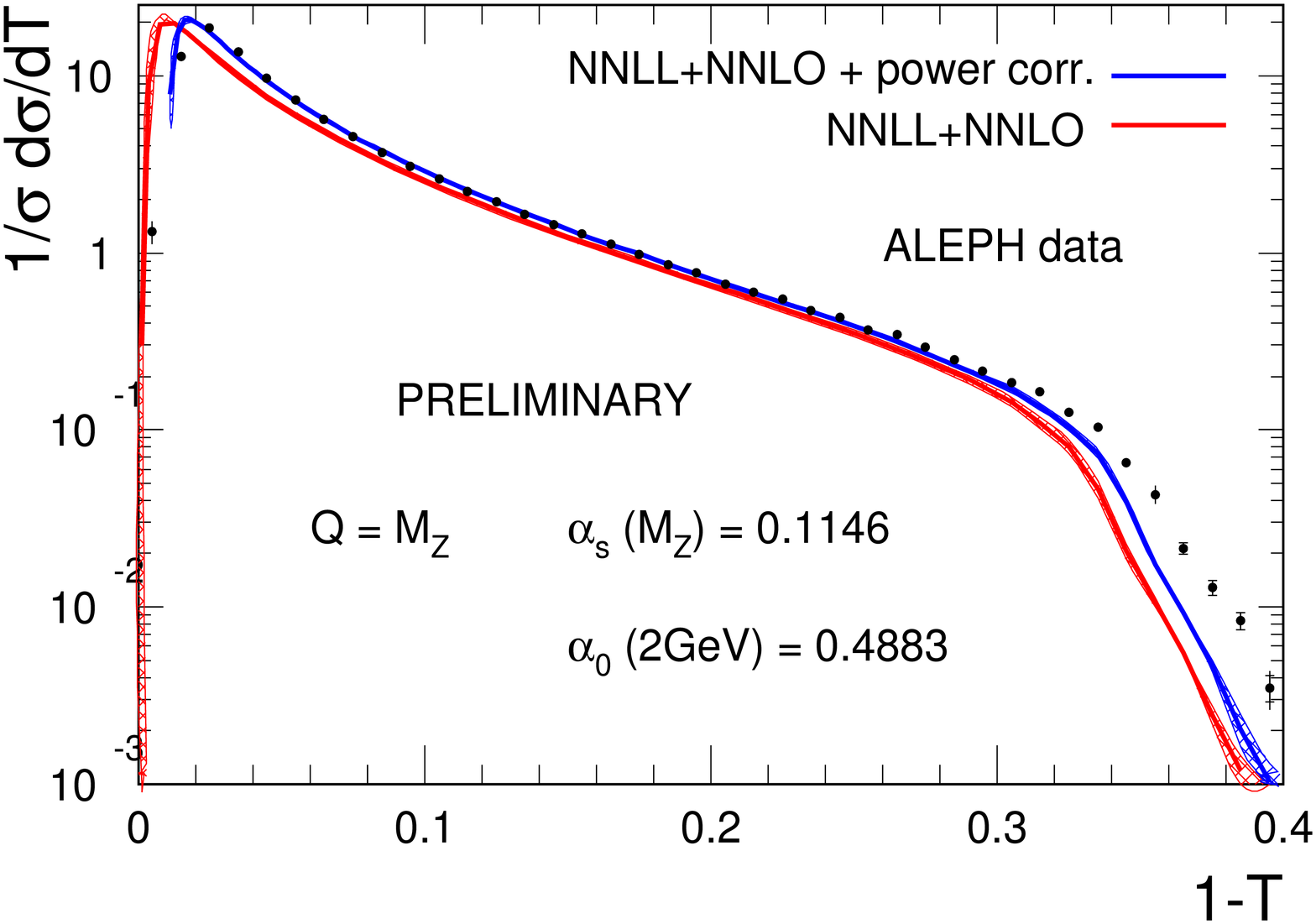}
    \caption{Theoretical prediction with (blue) and without (red)
	    power corrections compared to ALEPH data. The non-perturbative parameter $\alpha_{0}$ as well as
	    the strong coupling constant $\alpha_{s}$ are fitted to experimental data.}
    \label{fig:nonpt1}
 \end{minipage}
 \hspace{5mm}
 \begin{minipage}[c]{.47\textwidth}
    \centering
     \includegraphics[width=1.0\textwidth]{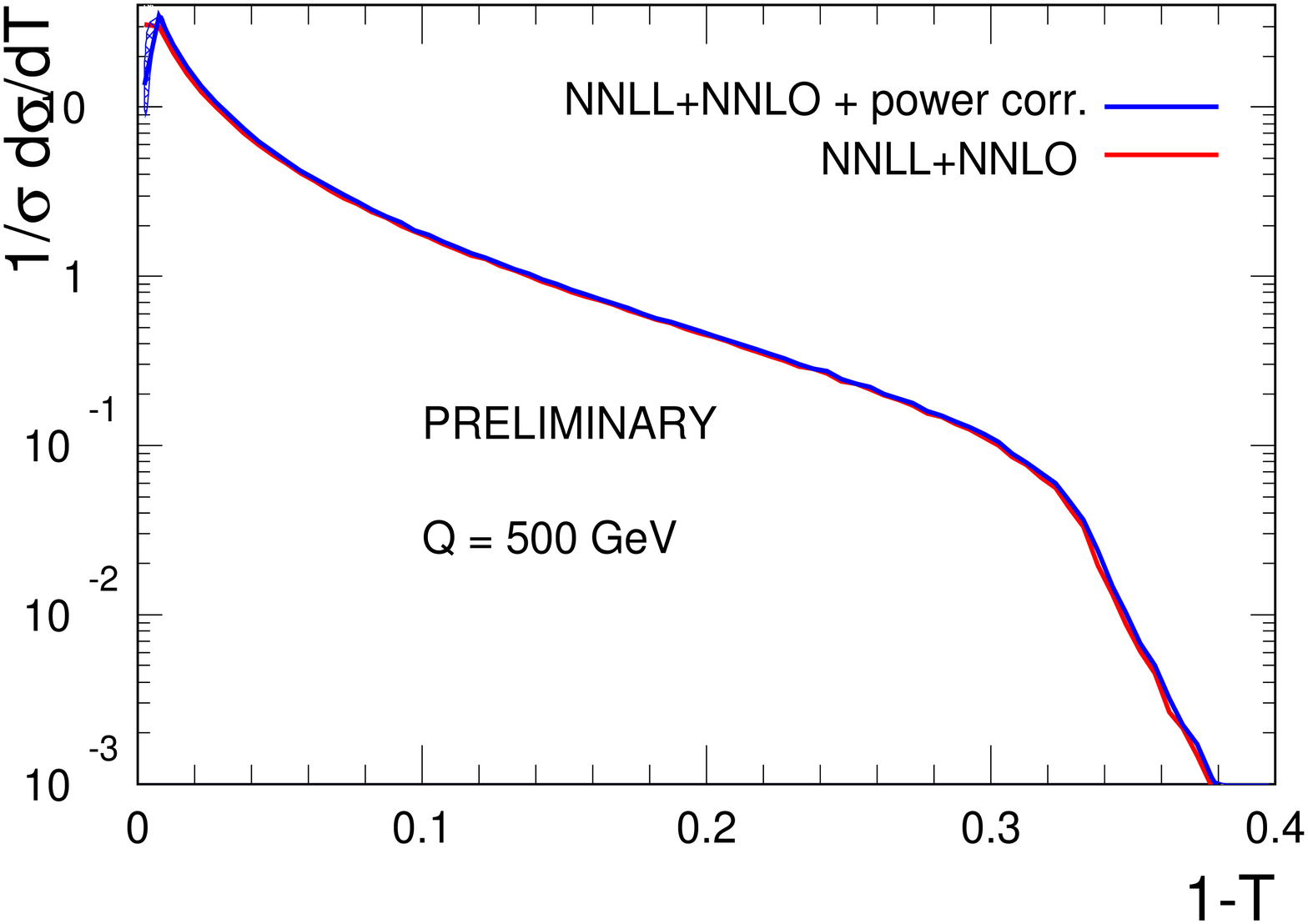}
    \caption{Comparison of the perturbative (red) and power-corrected (blue) distributions at a center of mass energy of 500 GeV.
	    The plot shows how much hadronisation corrections get suppressed at typical future linear collider energies.}
    \label{fig:nonpt2}
  \end{minipage}
\end{figure}

The recent results on event shape resummation improve the description of existing experimental data. In view of future work at
high energy linear colliders and precise determinations of the strong coupling constant, N$^2$LL predictions
for the remaining Event-Shape observables are necessary.
Moreover, an additional source of uncertainty is due to power-behaved hadronisation corrections which get large in the dijet region.
Currently there is no deep theoretical understanding of such corrections which constitute an important source of theoretical error.
In the past, these were often computed using leading-logarithmic parton shower Monte Carlo programs, which 
turned out to be clearly insufficient~\cite{Dissertori:2009ik} in view of the precision 
now attained by the perturbative description. Systematic approaches to 
hadronization within the dispersive model~\cite{alpha0,jaquier} or 
by using the shape function formalism~\cite{Abbate:2010xh,Korchemsky:1998ev} 
are offering  a more reliable description. 
Such corrections are quite sizeable at LEP energies (Fig. \ref{fig:nonpt1}) while they are highly suppressed at future linear colliders energies (Fig. \ref{fig:nonpt2}).
In Fig. \ref{fig:nonpt1} we show what the power-corrected distribution looks like when compared to the pure perturbative answer.
Non-perturbative corrections are computed with a dispersive model \cite{alpha0} and both the mean effective coupling $\alpha_{0}$ and the strong coupling $\alpha_{s}$ are
obtained by performing a simultaneous fit using ALEPH data at $Q=91.2$~GeV. Such a fit is purely qualitative since the correlation matrix is degenerate when only one data set is used.
To perform a meaningful fit, experimental data over a broader range of energies have to be included. We will address this issue in a future publication.

\section{Acknowledgments}

This research is supported in part by
the Swiss National Science Foundation (SNF) under contract
200020-138206, by the UK  STFC,  by the European Commission through the 
``LHCPhenoNet" Initial Training Network PITN-GA-2010-264564 and 
by the National Science Foundation under grant NSF PHY05-51164.


\begin{footnotesize}


\end{footnotesize}


\end{document}